\documentclass[twocolumn,amsmath,amssymb]{revtex4}
\usepackage[dvips]{graphicx}
\usepackage{dcolumn}
\usepackage{bm}

\begin{document}
\title[]
{Could some black holes have evolved from wormholes?}
\author{Peter K.\,F. Kuhfittig}
\address{Department of Mathematics\\
Milwaukee School of Engineering\\
Milwaukee, Wisconsin 53202-3109 USA}

\begin{abstract}\noindent
One way to explain the present acceleration of the Universe is
Einstein's cosmological constant.  It is quite likely, in view of
some recent studies, that a time-dependent equation of state had
caused the Universe to evolve from an earlier phantom-energy model.
In that case traversable wormholes could have formed spontaneously.
It is shown in this paper that such wormholes would eventually have
become black holes or quasi-black holes.  This would provide a
possible explanation for the huge number of black holes discovered,
while any evidence for the existence of wormholes is entirely lacking,
even though wormholes are just as good, in terms of being a prediction
of general relativity, as black holes.

\end{abstract}

\maketitle
\noindent
PAC numbers: 04.20.Jb, 04.20.Gz


\section{Introduction}\noindent
Traversable wormholes, whose possible existence was first conjectured
by Morris and Thorne in 1988 \cite{MT88}, are shortcuts or tunnels that
could in principle be used for traveling to remote parts of our
Universe or to different universes altogether.  Using units in which
$c=G=1,$ a wormhole can be described by the general line element
\begin{equation}\label{E:line1}
 ds^2=-e^{2\gamma(r)}dt^2+e^{2\alpha(r)}dr^2+r^2(d\theta^2
     +\text{sin}^2\theta\,d\phi^2).
\end{equation}
The motivation for this idea is the Schwarzschild line element
\begin{equation}\label{E:line2}
   ds^2=-\left(1-\frac{2M}{r}\right)dt^2+\frac{dr^2}{1-\frac{2M}{r}}
   +r^2(d\theta^2+\text{sin}^2\theta\,d\phi^2)
\end{equation}
which may be viewed as a black hole centered at the origin of a
$(t,r,\theta,\phi)$ coordinate system.  Both wormholes and black holes
are predictions of Einstein's general theory of relativity.  The main
difference is that wormholes must necessarily violate certain energy
conditions; more precisely, the stress-energy tensor of the matter
source of gravity violates the weak energy condition \cite{MT88}.  If
the matter source was different in the past, then wormholes could
theoretically have been formed naturally.  Examples are equations of
state that parameterize certain dark energy models, as we will see in
the next section. Moreover, according to some recent studies, such a
transition is likely to have occurred in the relatively recent past
\cite{SS1, SS2}.

In this paper we study an equation of state that is both space and
time dependent.  It is proposed that if the equation of state evolved
into the present cosmological constant model, then any wormhole that
had previously come into existence would have formed an event horizon,
thereby becoming a black hole or a quasi-black hole..  This would
provide a possible explanation for the failure to detect any evidence
of wormholes, while black holes appear to be abundant.

A unified view of wormholes and black holes, including possible
interconversions, was first proposed by Hayward \cite{sH99}.  For a
detailed discussion and analysis, see \cite{KH04} and references
therein.  (A highly advanced civilization might also be able
to reverse the natural process discussed in this paper to create
a traversable wormhole.)

\section{Background}\label{S:background}\noindent
Interest in traversable wormholes has increased in recent years due
to an unexpected connection, the discovery that our Universe is
undergoing an accelerated expansion \cite{aR98, sP99}.  This
acceleration, caused by a negative pressure \emph{dark energy},
implies that $\overset{..}{a}>0$ in the Friedmann equation
$\frac{\overset{..}{a}}{a}=-\frac{4\pi}{3}(\rho+3p).$  The equation
of state is $p=-K\rho$, $K>\frac{1}{3}$, and $\rho>0.$  While the
condition $K>\frac{1}{3}$ is required for an accelerated expansion,
larger values for $K$ are also of interest.  In fact, the most
appealing candidate for dark energy is $K=1$, corresponding to
the cosmological constant $\Lambda$ \cite{mC01}.  The presence of the
cosmological constant has resulted in a modification of the Einstein
field equations by effectively adding an isotropic and homogeneous
source with constant equation of state \cite{jA06}.  One can
therefore argue, as in Ref.~\cite{rB07}, that this model is the
primary candidate for the present Universe.

Another widely studied possibility is the case $K>1$, referred to as
\emph{phantom energy} \cite{rC02}  To see why, we need to recall that
the set of \emph{orthonormal basis vectors} may be interpreted as the
proper frame of a set of observers who remain at rest in the coordinate
system.  If the basis vectors are denoted by
$e_t$, $e_r$, $e_{\theta}$, and $e_{\phi}$, then the orthonormal
basis vectors are (referring to line element (\ref{E:line1})),
\begin{align*}
    e_{\hat{t}}=&\,\,e^{-\gamma(r)}e_t,\qquad
    e_{\hat{r}}=e^{-\alpha(r)}e_r,\\
    e_{\hat{\theta}}=&\,\,r^{-1}e_{\theta},\qquad\quad
    e_{\hat{\phi}}=(r\,\text{sin}\,\,\theta)^{-1}e_{\phi}.
\end{align*}
In this frame of reference the components of the stress-energy tensor
$T_{\hat{\alpha}\hat{\beta}}$ have an immediate physical
interpretation: $T_{\hat{t}\hat{t}}=\rho$, $T_{\hat{r}\hat{r}}=p$,
$T_{\hat{\theta}\hat{\theta}}=T_{\hat{\phi}\hat{\phi}}=p_t$, where
$\rho$ is the energy density, $p$ the radial pressure, and $p_t$ the
lateral pressure.  The weak energy condition (WEC) can now be stated as
follows: $T_{\hat{\alpha}\hat{\beta}}\mu^{\hat{\alpha}}
\mu^{\hat{\beta}}\ge 0$ for all time-like vectors and, by continuity,
all null vectors.  For example, given the radial outgoing null vector
$(1,1,0,0)$, we obtain
\[
    T_{\hat{\alpha}\hat{\beta}}\mu^{\hat{\alpha}}\mu^{\hat{\beta}}
      =\rho+p\ge 0.
\]
So if the WEC is violated, we have $\rho+p<0.$  While all classical
forms of matter ordinarily meet this condition, there are situations
in quantum field theory, such as the Casimir effect, that allow such
violations \cite{MTY88}.  More importantly, in our case this violation
occurs whenever $K>1$, thereby meeting the primary prerequisite
for the existence of wormholes.

Two recent papers \cite{fL07, fR07} discuss
wormhole solutions that depend on a variable equation of state
parameter, i. e., $\frac{p}{\rho}=-K(r),$ where $K(r)>1$ for all $r$.
The variable $r$ refers to the radial coordinate in the line element
\begin{equation}\label{E:line3}
  ds^2=-e^{2h(r)}dt^2+\frac{1}{1-\frac{b(r)}{r}}dr^2
       +r^2(d\theta^2+\text{sin}^2\theta\,d\phi^2).
\end{equation}
In this form of the line element, $h=h(r)$ is called the \emph
{redshift function} and $b=b(r)$ the \emph{shape function}.  The
minimum radius $r=r_0$ corresponds to the \emph{throat} of the
wormhole, where $b(r_0)=r_0.$

It is shown in Ref. \cite{fR07} that given a specific shape function,
it is possible to determine $K(r)$ and vice versa.  It is also
assumed that $h'(r)\equiv 0$, referred to as the ``zero
tidal-force solution" in Ref.~\cite{MT88}.

An earlier study \cite{BS94} assumed that the equation of state is
time dependent.  In this paper we will assume that the equation
depends on both $r$ and $t$, while remaining independent of direction.

Since the equation of state is time dependent, we assume that the
corresponding metric is also time dependent.  It is shown in the next
section that such a metric describes a slowly evolving wormhole
structure without assigning specific functions to $h$, $b$,
and $K$.  In particular, the function $h$ in line element
(\ref{E:line3}) need not be a constant.

Evolving wormhole geometries are also discussed in Refs.
\cite{lA98} and \cite{KS96}.

Since we are dealing with a given time-dependent equation of state,
it is natural to consider the consequences of an evolving equation,
particularly one in which the parameter $K(r,t)$ approaches unity.
That is the topic of Sec. \ref{S:evolving}.

\section{The metric}\label{S:metric}\noindent
In this paper we will be dealing with a time-dependent metric
describing an evolving wormhole:
\begin{equation}\label{E:line4}
   ds^2 =-e^{2\gamma(r,t)}dt^2+e^{2\alpha(r,t)}dr^2+r^2(d\theta^2+
      \text{sin}^2\theta\, d\phi^2),
\end{equation}
where $\gamma$ and $\alpha$ have continuous partial derivatives, so
that $\gamma$ and $\alpha$ are continuous, as well.

In view of line element (\ref{E:line3}), we have
\begin{equation}\label{E:alpha}
   e^{2\alpha(r,t)}=\frac{1}{1-b(r,t)/r}.
\end{equation}
So the shape function is now given by
\begin{equation}\label{E:shape1}
    b(r,t)=r(1-e^{-2\alpha(r,t)}).
\end{equation}
To study the effect of a gradually changing equation of state, we
assume the existence of a fixed throat at $r=r_0$, that is,
$b(r_0,t)=r_0$ for all $t$.  (In other words, the wormhole is close
to being static for relatively long periods of time.)  As a
consequence, for all $t$,
\[
   \lim_{r \to r_0}\alpha(r,t)=+\infty \quad \text{and} \quad
   \lim_{r \to r_0}\frac{\partial}{\partial r}\alpha(r,t)=-\infty.
\]
As before, $\gamma=\gamma(r,t)$ is the redshift function, which
must be everywhere finite to prevent an event horizon at the throat.
As in the case of the Schwarzschild line element,
$\frac{\partial}{\partial r}\gamma(r,t)>0.$  The qualitative features of
$\alpha$ and $\gamma$ are shown in Fig. 1.

To obtain a traversable wormhole, the shape function must obey the
usual flare-out conditions at the throat, modified to accommodate the
time dependence:
\[
  b(r_0,t)=r_0\quad \text{and} \quad \frac{\partial}{\partial r}
       b(r_0,t)<1 \quad \text{for all}\,\,t.
\]
So $b(r,t)<r$ for all $t$ near the throat.  Another requirement
is asymptotic flatness: $b(r,t)/r\rightarrow 0$ as
$r\rightarrow\infty.$

The next step is to list the time-dependent components of the
Einstein tensor in the orthonormal frame.  (For a derivation, see
Kuhfittig \cite{pK02}.)
\begin{equation}\label{E:Einstein1}
   G_{\hat{t}\hat{t}}=\frac{2}{r}e^{-2\alpha(r,t)}\frac{\partial}
   {\partial r}\alpha(r,t)+\frac{1}{r^2}(1-e^{-2\alpha(r,t)}),
\end{equation}
\begin{equation}\label{E:Einstein2}
    G_{\hat{r}\hat{r}}=\frac{2}{r}e^{-2\alpha(r,t)}\frac{\partial}
   {\partial r}\gamma(r,t)-\frac{1}{r^2}(1-e^{-2\alpha(r,t)}),
\end{equation}
\begin{equation}\label{E:Einstein3}
   G_{\hat{t}\hat{r}}=\frac{2}{r}e^{-\gamma(r,t)}
    e^{-\alpha(r,t)}\frac{\partial}{\partial t}\alpha(r,t),
\end{equation}
\begin{multline}\label{E:Einstein4}
   G_{\hat{\theta}\hat{\theta}}=G_{\hat{\phi}\hat{\phi}}=
   -e^{-2\gamma(r,t)}\left[\frac{\partial^2}{\partial t^2}
    \alpha(r,t)
     \phantom{\left(\frac{\partial}{\partial t}\alpha(r,t)
    \right)^2}\right.\\
    \left.-\frac{\partial}{\partial t}\gamma(r,t)
    \frac{\partial}{\partial t}\alpha(r,t)
      \right.
     \left.+\left(\frac{\partial}{\partial t}\alpha(r,t)
    \right)^2\right]\\
     -e^{-2\alpha(r,t)}\left[-\frac{\partial^2}{\partial r^2}
    \gamma(r,t)
     \phantom{\left(\frac{\partial}{\partial t}\alpha(r,t)
    \right)^2}\right.\\
    \left.+\frac{\partial}{\partial r}\gamma(r,t)
    \frac{\partial}{\partial r}\alpha(r,t)
      \right.
     \left.-\left(\frac{\partial}{\partial r}\gamma(r,t)
    \right)^2\right]\\
  -\frac{1}{r}e^{-2\alpha(r,t)}\left(-\frac{\partial}{\partial r}
   \gamma(r,t)+\frac{\partial}{\partial r}\alpha(r,t)\right).
\end{multline}

Recall that from the Einstein field equations in the orthonormal frame,
$G_{\hat{\alpha}\hat{\beta}}=8\pi T_{\hat{\alpha}\hat{\beta}}$,
the components of the Einstein tensor are proportional to the
components of the stress-energy tensor.  In particular,
$T_{\hat{t}\hat{t}}=\rho(r,t)$, $T_{\hat{r}\hat{r}}=p(r,t)$,
$T_{\hat{\theta}\hat{\theta}}=T_{\hat{\phi}\hat{\phi}}=
p_t(r,t),$ and $T_{\hat{t}\hat{r}}=T_{\hat{r}\hat{t}}=\frac{1}{8\pi}
G_{\hat{t}\hat{r}}=g(r,t)$, where $f(r,t)=-g(r,t)$ is usually
interpreted as the energy flux in the outward radial direction
\cite{tR93}.  For the outgoing null vector $(1,1,0,0)$, the WEC,
$T_{\hat{\alpha}\hat{\beta}}\mu^{\hat{\alpha}}
\mu^{\hat{\beta}}\ge 0$, now becomes $\rho+p\pm 2g\ge 0.$

\section{The equation of state}\noindent
Since Eq.~(\ref{E:line4}) describes a slowly evolving wormhole, the
equation of state should have a time-dependent parameter $K(r,t)$.
As in Ref.~\cite{fR07}, $K$ depends on the radial coordinate, but
not on the direction.  To be compatible with the wormhole geometry
in Sec.~\ref{S:metric}, Francisco Lobo suggested the inclusion
of a term analogous to the flux term \cite{fL08}.  One such
possibility is
\begin{equation}\label{E:eqstate1}
   p(r,t)=-K(r,t)[\rho(r,t)+2g(r,t)],
\end{equation}
where $|2g(r,t)|<\rho(r,t).$  The last condition implies that
$\alpha(r,t)$ changes slowly enough; in fact, $g(r,t)$ can be
identically zero.  If $K(r,t)>1,$ this equation of state describes
a generalized phantom-energy model, as we will see below
[Eq.~(\ref{E:WEC})].  As a result, the case $K=1$ would still
correspond to a cosmological constant.

Since the notion of dark or
phantom energy applies only to a homogeneous distribution of
matter in the Universe, while wormhole spacetimes are necessarily
inhomogeneous, we adopt the point of view in Sushkov \cite{sS05}:
extended to spherically symmetric wormhole geometries, the pressure
appearing in the equation of state is now a negative radial pressure,
while the transverse pressure is determined from the field equations.

Given the evolving equation of state (\ref{E:eqstate1}), supppose
at some point in the past, the equation of state parameter $K$ had
actually crossed the phantom divide, so that, for a time,
$K(r,t)>1$. In fact, according to Refs. \cite{SS1, SS2}, it is
quite likely that such a transition had taken place in the relatively
recent past.  Suppose further that $K$ was decreasing, i.e.,
$\frac{\partial}{\partial t}K(r,t)<0$ with $K(r,t)\rightarrow 1$ at a
time closer to the present, and that $K$ decreased fast enough
during this time interval to compensate for the very gradual
increase in the energy density characteristic of phantom energy,
allowing us to assume that
\begin{equation}\label{E:pospressure}
    \frac{\partial}{\partial t}p(r,t)\ge 0.
\end{equation}
With this information we can now determine
the sign of $g(r,t)$:
\begin{multline*}
   \frac{\partial}{\partial t}p(r,t)=\frac{1}{8\pi}
   \frac{\partial G_{\hat{r}\hat{r}}}{\partial t}\\
  =\frac{1}{8\pi}e^{-2\alpha(r,t)}\left[\frac{\partial}{\partial t}
   \alpha(r,t)\left((-2)\frac{2}{r}\frac{\partial}
    {\partial r}\gamma(r,t)-\frac{2}{r^2}\right)\right.\\
    \left.+\frac{2}{r}\frac{\partial}{\partial t}
    \left(\frac{\partial}{\partial r}\gamma(r,t)\right)\right]\ge 0.
\end{multline*}
Since the pressure would change at a finite rate, it follows that
$\frac{\partial}{\partial t}\alpha(r,t)$ is finite.  Also,
\begin{equation}\label{E:derivalpha}
  \frac{\partial}{\partial t}\alpha(r,t)\le 0
   \quad \text{for all}\quad r,
\end{equation}
so that $g(r,t)\le 0$, as well.  The reason is that, judging from Fig.~1,
the time rate of change of the slope of $\gamma$, that is,
\[
  \frac{\partial}{\partial t}\left(\frac{\partial}{\partial r}
      \gamma(r,t)\right),
\]
is likely to be vanishingly small.  We will return to this point in
Sec. \ref{S:evolving}.

\begin{figure}[ptb]
\begin{center}
\includegraphics[width=0.5\textwidth]{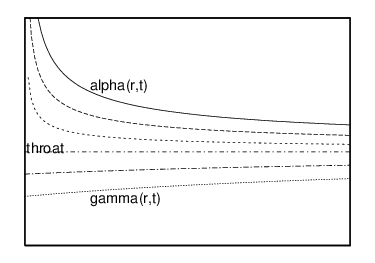}
\end{center}
\caption{Graph showing the qualitative features of
$\alpha(r,t)$ and $\gamma(r,t)$.}
\end{figure}

Returning to Eq.~(\ref{E:eqstate1}), observe that
\begin{multline}\label{E:WEC}
   T_{\hat{\alpha}\hat{\beta}}\mu^{\hat{\alpha}}
   \mu^{\hat{\beta}}=\rho+p+2g\\=\rho(r,t)-K(r,t)[\rho(r,t)+2g(r,t)]+2g(r,t)\\
    =[\rho(r,t)+2g(r,t)][-K(r,t)+1]<0
\end{multline}
since $K(r,t)>1$ and $|2g(r,t)|<\rho(r,t)$.  So the WEC is violated, as
one would expect in a phantom-energy scenario.  We would also expect
the flare-out conditions to be met.  That is the topic of the next
section.


\section{The flare-out conditions}\noindent
From the Einstein field equations $G_{\hat{\alpha}\hat{\beta}}=
8\pi T_{\hat{\alpha}\hat{\beta}}$ and the above equation of state,
we have
\[
  G_{\hat{t}\hat{t}}=8\pi\rho\quad \text{and}\quad
  G_{\hat{r}\hat{r}}=8\pi[-K(r,t)][\rho(r,t)+2g(r,t)].
\]
Using Eqs.~(\ref{E:Einstein1})-(\ref{E:Einstein3}), we obtain the
following system of equations:
\begin{equation*}
  G_{\hat{t}\hat{t}}=8\pi T_{\hat{t}\hat{t}}=\frac{2}{r}
   e^{-2\alpha(r,t)}\frac{\partial}{\partial r}\alpha(r,t)
   +\frac{1}{r^2}\left(1-e^{-2\alpha(r,t)}\right)
\end{equation*}
and
\begin{multline*}
  G_{\hat{r}\hat{r}}=8\pi T_{\hat{r}\hat{r}}=8\pi[-K(r,t)]
     [\rho(r,t)+2g(r,t)]\\
    =\frac{2}{r}e^{-2\alpha(r,t)}\frac{\partial}{\partial r}
    \gamma(r,t)-\frac{1}{r^2}\left(1-e^{-2\alpha(r,t)}\right).
\end{multline*}
Substituting the expressions for $\rho(r,t)$ and $2g(r,t)$ yields
\begin{multline}\label{E:equation1}
  -K(r,t)\frac{2}{r}e^{-2\alpha(r,t)}\frac{\partial}{\partial r}
  \alpha(r,t)\\
   =K(r,t)\frac{1}{r^2}\left(1-e^{-2\alpha(r,t)}\right)+\frac{2}{r}
     e^{-2\alpha(r,t)}\frac{\partial}{\partial r}\gamma(r,t)\\
        -\frac{1}{r^2}\left(1-e^{-2\alpha(r,t)}\right)\\
   +K(r,t)\frac{4}{r}e^{-\gamma(r,t)}e^{-\alpha(r,t)}
          \frac{\partial}{\partial t}\alpha(r,t).
\end{multline}
The following rearrangement will be needed again in Sec.~
\ref{S:evolving}:
\begin{multline}\label{E:equation2}
   \frac{\partial}{\partial r}\alpha(r,t)+2e^{-\gamma(r,t)}
      e^{\alpha(r,t)}\frac{\partial}{\partial t}\alpha(r,t)\\
   =-\frac{1}{2r}\left(e^{2\alpha(r,t)}-1\right)+\frac{1}{2r}
       \frac{1}{K(r,t)}\left(e^{2\alpha(r,t)}-1\right)\\
     -\frac{1}{K(r,t)}\frac{\partial}{\partial r}\gamma(r,t).
\end{multline}
For the purpose of analysis, however, a more convenient form is
the following:
\begin{multline}\label{E:equation3}
  \frac{2\frac{\partial}{\partial r}\alpha(r,t)}
      {e^{2\alpha(r,t)}-1}=\frac{-2\frac{\partial}{\partial r}
         \gamma(r,t)}{K(r,t)\left(e^{2\alpha(r,t)}-1\right)}\\
   -\frac{1}{r}\left(1-\frac{1}{K(r,t)}\right)-
      \frac{4e^{-\gamma(r,t)}e^{\alpha(r,t)}\frac{\partial}
               {\partial t}\alpha(r,t)}
    {e^{2\alpha(r,t)}-1}.
\end{multline}
At this point we define
\begin{multline}\label{E:F}
  F(r,t)=\frac{-2\frac{\partial}{\partial r}
         \gamma(r,t)}{K(r,t)\left(e^{2\alpha(r,t)}-1\right)}\\
   -\frac{1}{r}\left(1-\frac{1}{K(r,t)}\right)-
      \frac{4e^{-\gamma(r,t)}e^{\alpha(r,t)}\frac{\partial}
               {\partial t}\alpha(r,t)}
    {e^{2\alpha(r,t)}-1}.
\end{multline}
The form $\int\frac{du}{e^u-1}=\text{ln}\,(e^u-1)-u$ now yields
\[
  \text{ln}\left(e^{2\alpha(r,t)}-1\right)-2\alpha(r,t)=
    \int^r_cF(r',t)dr',
\]
where $c$ is an arbitrary constant.  Recalling that $\frac{\partial}
{\partial t}\alpha(r,t)$ is finite, observe that for any fixed $t$,
$F(r,t)$ is sectionally continuous for $r\ge r_0$, so that the
integral exists for $c\ge r_0.$  Thus we may write
\[
   e^{2\alpha(r,t)}-1=e^{2\alpha(r,t)+\int^r_cF(r',t)dr'},
\]
whence
\begin{equation}\label{E:shape2}
   1-e^{-2\alpha(r,t)}=e^{\int^r_cF(r',t)dr'}.
\end{equation}
From Eq.~(\ref{E:shape1}) we have $b(r,t)=re^{\int^r_cF(r',t)dr'}.$
So the requirement $b(r_0,t)=r_0$ now determines the arbitrary
constant: $c=r_0.$ Thus
\[
   b(r,t)=re^{\int^r_{r_0}F(r',t)dr'}.
\]
Differentiating, we get for $r=r_0$
\begin{multline*}
  \frac{\partial}{\partial r}b(r_0,t)=e^{\int^{r_0}_{r_0}F(r,t)dr}+
   r_0e^{\int^{r_0}_{r_0}F(r,t)dr}\\
  \times\left[\frac{-2\frac{\partial}{\partial r}\gamma(r_0,t)}
   {K(r_0,t)\left(e^{2\alpha(r_0,t)}-1\right)}-\frac{1}{r_0}
    \left(1-\frac{1}{K(r_0,t)}\right)\right.\\
    \left.-\frac{4e^{-\gamma(r_0,t)}e^{\alpha(r_0,t)}
        \frac{\partial}{\partial t}\alpha(r_0,t)}
           {e^{2\alpha(r_0,t)}-1}\right]\\
    =\frac{1}{K(r_0,t)}<1
\end{multline*}
by the assumption $K(r,t)>1$.

The line element now becomes
\begin{multline*}
  ds^2=-e^{2\gamma(r,t)}dt^2+\frac{dr^2}{1-e^{\int^r_{r_0}F(r',t)dr'}}\\
   +r^2(d\theta^2+\text{sin}^2\theta\,d\phi^2).
\end{multline*}
Since the flare-out conditions have been satisfied, the line element
describes a traversable wormhole.

As an illustration, if $\alpha$ is time-independent,
$\gamma(r,t)$ is a constant, and $K(r,t)=K$ is also a constant,
then it follows from Eq.(\ref{E:F}) that
\begin{equation*}
   e^{\int^r_{r_0}F(r',t)dr'}=e^{\int^r_{r_0}
     \left[-\frac{1}{r'}\left(1-\frac{1}{K}\right)\right]dr'}
        =\left(\frac{r_0}{r}\right)^{1-1/K}
\end{equation*}
and
\[
   e^{2\alpha(r)}=\frac{1}{1-\left(\frac{r_0}{r}\right)^{1-1/K}},
\]
which is Lobo's solution \cite{fL05}.

At this point the following remark is in order: since we are only interested in
the possible existence of wormholes, it is sufficient to note that to complete
the description, the wormhole material should be cut off at some $r=a$ and joined
to an external Schwarzschild spacetime.  (See Refs. \cite{fL05,oZ05,pK06} for
details.)  This junction will make the space asymptotically flat, a critical
feature referred to in the next section.


\section{Implications}\label{S:evolving}\noindent
As we have seen, since $K(r,t)>1$, line element~(\ref{E:line4}) describes a
traversable wormhole as long as the qualitative features in Fig. 1 are met,
resulting in a violation of the WEC.  It is therefore conceivable that wormholes
had formed spontaneously during the phantom-energy phase.  Moreover, a possible
mechanism for the formation of such wormholes is discussed in Ref.~\cite{pG04}.

In this section we study the consequences of our assumption that $K(r,t)
\rightarrow 1$ sufficiently fast some time in the past.  (As noted
earlier, this limiting case corresponds to a cosmological constant.)
Before doing so, however, let us recall that for an ``arbitrary"
wormhole the rate of change of the slope of $\gamma$ is likely to be vanishingly
small, leading to inequality (\ref{E:derivalpha}).
In addition,

\begin{equation*}
   \lim_{r \to r_0}\alpha(r,t)=+\infty\quad \text{and}\quad
   \lim_{r \to r_0}\frac{\partial}{\partial r}\alpha(r,t)=-\infty,
\end{equation*}
for all $t$, while $\gamma=\gamma(r_0,t)$ must be finite
to prevent an event horizon.  Also, $\lim_{r \to\infty}\alpha(r,t)=
\lim_{r \to\infty}\gamma(r,t)=0$ for all $t$.  Finally,
$\frac{\partial}{\partial r}\gamma(r,t)>0.$  (See Fig. 1.)

In Eq.~(\ref{E:equation2}), as $K(r,t)\rightarrow 1$, we are left with
\begin{equation*}
  \frac{\partial}{\partial r}\alpha(r,t)+2e^{-\gamma(r,t)}e^{\alpha(r,t)}
   \frac{\partial}{\partial t}\alpha(r,t)
      =-\frac{\partial}{\partial r}\gamma(r,t).
\end{equation*}
As $r\rightarrow r_0$, the left side goes to $-\infty$ (since
$\frac{\partial}{\partial t}\alpha(r,t)$ is finite and nonpositive), so that
the right side yields
\begin{equation}\label{E:eventhorizon}
   \lim_{r\to r_0}\frac{\partial}{\partial r}\gamma(r,t)=+\infty.
\end{equation}
Eq.~(\ref{E:eventhorizon}) implies that either (i) $\gamma=\gamma(r,t)$
must approach the vertical line $r=r_0$ asymptotically, i.e.,
$\lim_{r\to r_0}\gamma(r,t)=-\infty,$ or (ii)
$\frac{\partial}{\partial r}\gamma(r,t)$ is undefined at $r=r_0$, i.e.,
$\gamma$ has a vertical tangent at $(r_0, \gamma(r_0,t)$).  The more
plausible case (i) leads to an event horizon at the throat.  At first
glance the outcome is a black hole, since,
in asymptotically flat spacetimes, a black hole is essentially characterized
by the impossibility of escaping from a certain prescribed region to future
null infinity.  More formally, the boundary of $J^{-1}(\mathcal{I}^{+})$, the
causal past of future null infinity, must be a region that does not include the
entire spacetime.  But this requirement is met because our starting point is
a well-defined region, the throat of a wormhole.

A possible problem with this conclusion is that case (ii) needs to be
examined more closely: according to Sushkov and Zaslavskii \cite{SZ09},
the would-be event horizon may be singular, having infinite surface
stresses.  To see this, let us rewrite the metric, Eq. (\ref{E:line4}),
in the form
\begin{multline}\label{E:line5}
  ds^2=-e^{2\gamma_{\pm}(r,t)}dt^2+\frac{dr^2}{1-b_{\pm}(r,t)/r}\\
       +r^2(d\theta^2+\text{sin}^2\theta\,d\phi^2)
\end{multline}
and introduce the proper radial distance
\[
    \ell(r)=\pm\int^r_{r_0}\frac{dr'}{\sqrt{1-b_{\pm}(r',t)/r'}}.
\]
Then Eq. (\ref{E:line5}) becomes
\begin{equation}\label{E:line6}
   ds^2=-e^{2\gamma(r(\ell),t)}dt^2+d\ell^2+r^2(\ell)
       (d\theta^2+\text{sin}^2\theta\,d\phi^2),
\end{equation}
where $-\infty<\ell<\infty$ and the throat is at $\ell=0$.  Now
$\lim_{r \to r_0}\gamma_{\pm}(r,t)$ is equivalent to $\lim_{\ell\to 0}
N(\ell,t)$ for the lapse function $N$.  In Ref. \cite{SZ09},
the lapse function $N$ corresponds to our $e^{\gamma(r(\ell),t)}$
but is actually a time-independent function.  The Lanczos
equations now yield the following expression for the surface
stresses at the throat:
\begin{equation}\label{E:Lanczos}
    8\pi S^0_{\phantom{0}0}
    =-[K^0_{\phantom{0}0}]=\frac{2}{N}
    \left[\left(\frac{\partial N}{\partial\ell}\right)_{+}-
    \left(\frac{\partial N}{\partial\ell}\right)_{-}\right],
    \end{equation}
where the $+$ and $-$ refer to the right- and left-hand limits,
    respectively, at $\ell=0$.  Observe next that
\begin{equation}
  \frac{\partial\gamma(r,t)}{\partial\ell}=
     \frac{\partial\gamma(r,t)}{\partial r}
     \frac{\partial r}{\partial\ell}=\sqrt{1-\frac{b(r,t)}{r}}
       \frac{\partial\gamma(r,t)}{\partial r}.
\end{equation}
We already know that
$\lim_{r\to r_0}\frac{\partial}{\partial r}\gamma_{\pm}(r,t)=+\infty$.  But
suppose that $\frac{\partial}{\partial r}\gamma_{\pm}(r,t)$ does not diverge
any faster than $[1-b_{\pm}(r,t)/r]^{-1/2}$.  Then
\begin{equation*}
  \left(\frac{\partial N}{\partial\ell}\right)_{\pm}=
    \left(e^{\gamma(r,t)}\sqrt{1-\frac{b(r,t)}{r}}
       \frac{\partial\gamma(r,t)}{\partial r}\right)_{\pm}=0
\end{equation*}
and
\begin{equation*}
   \left(\frac{\partial N}{\partial\ell}\right)_{+}-
  \left(\frac{\partial N}{\partial\ell}\right)_{-}= 0.
\end{equation*}
This case corresponds to Situation TB (case 1 with $\varepsilon=0$) in
Ref. \cite{SZ09}, leading to a finite Kretschmann scalar, since
$K(r,t)\rightarrow 1$ is equivalent to $\ell\rightarrow 0$.  The
difference in our situation is that  $S^0_{\phantom{0}0}$ is now equal
to $0$, as can be seen from Eq. (\ref{E:Lanczos}):
\begin{multline*}
   4\pi S^0_{\phantom{0}0}=\left[
   \left(e^{\gamma(r,t)}\right)_{+}
   \left(\sqrt{1-\frac{b(r,t)}{r}}\frac{\partial\gamma(r,t)}
      {\partial r}\right)_{+}-\right.\\
      \left.
      \left(e^{\gamma(r,t)}\right)_{-}
   \left(\sqrt{1-\frac{b(r,t)}{r}}\frac{\partial\gamma(r,t)}
      {\partial r}\right)_{-}\right]/ e^{\gamma(r,t)}\\=0
\end{multline*}
by the continuity of $e^{\gamma_{\pm}(r,t)}$ [in particular,
$\gamma_{+}(r_0,t)=\gamma_{-}(r_0,t)$] for both (i) and (ii).  Case
(i) leads to a black hole, as we saw earlier.  In case (ii), we
are still dealing with a wormhole, but this case must be
excluded for physical reasons since the gradient of the
redshift function is related to the tidal constraints, so that
$\frac{\partial}{\partial r}\gamma(r_0,t)$ would have remained
finite.  With case (ii) eliminated, the end result is a black hole.

If $\frac{\partial}{\partial r}\gamma(r,t)$ diverges too rapidly,
however, then $S^0_{\phantom{0}0}$ may no longer be finite.  The result
is a quasi-black hole.  (See Ref. \cite{LZ08} for a discussion.)  For
example, if $\gamma(r,t)=(1/4)\text{ln}[1-b(r,t)/r]$ (omitting the
subscript $\pm$), then
\begin{equation*}
  \frac{\partial N}{\partial\ell}=
     \frac{b(r,t)-r\partial b(r,t)/\partial r}{4r^2}
          \frac{1}{[1-b(r,t)/r]^{1/4}},
\end{equation*}
which diverges at the throat.

These conclusions are not in conflict with those in Ref. \cite{SZ09} since
our wormholes are non-generic, being spherically symmetric and non-static.
More importantly, we are making a number of additional assumptions: the
qualitative features in Fig. 1 are to be met, $K(r,t)$ increases rapidly
enough, the time rate of change of $\frac{\partial}{\partial r}\gamma(r,t)$
is vanishingly small, and $\frac{\partial}{\partial r}\gamma(r,t)$ diverges
sufficiently slowly at the throat.

Quasi-black holes possess quasihorizons, rather than event horizons.  Since
our quasi-black holes started off as wormholes, they are technically still
wormholes.  Having infinite surface stresses, they would not be traversable
by humanoid travelers, but they might still be capable of transmitting
signals.

In summary, this paper discusses a wormhole geometry supported by a generalized
form of phantom energy: the equation of state is
$p(r,t)=-K(r,t)[\rho(r,t)+2g(r,t)]$, where $K(r,t)>1$ for all $t$.  It is
quite likely that the equation of state had crossed the phantom divide in the
relatively recent past.  If, in addition, the equation of state
evolved so that $K(r,t)\rightarrow 1$ closer to the present,
the result is a spacetime with a cosmological constant, a primary candidate
for a model of the present Universe.  It is shown that any existing wormhole
would form an event horizon or a quasihorizon at the throat, resulting either
in a black hole or a quasi-black hole.  Assuming that wormholes could have
formed naturally during the phantom-energy phase, this outcome provides at
least a \emph{possible} explanation for the abundance of black holes and
the complete lack of wormholes, even though wormholes are just as good a
prediction of general relativity as black holes.

\section{Additional comment}\noindent
The evolution of wormholes discussed in this paper suggests
a relatively simple way for a highly advanced civilization to construct a
wormhole: identify a black hole that has evolved, submerge the black hole in
phantom dark energy, and reverse the process.  (For a general discussion of
wormhole construction from black holes using phantom energy, see Ref.
\cite{KH04}.)

\end{document}